\documentclass[twocolumn,aps,prb,showpacs]{revtex4}
\usepackage{amssymb}
\usepackage{amsmath}
\usepackage{graphicx}
\usepackage{epsfig}
\newcommand{\be}{\begin{equation}}
\newcommand{\ee}{\end{equation}}
\newcommand{\bes}{\begin{equation}\begin{split}}
\newcommand{\ese}{\end{split}\end{\equation}}
\newcommand{\bea}{\begin{eqnarray}}
\newcommand{\eea}{\end{eqnarray}}
\def\n{\noindent}
\def\etal{{\sl et al.}}

\begin{document}

\title{Theory of a double-dot charge detector}
\author{Tam\'as Geszti and J\'ozsef Zsolt Bern\'ad}
\affiliation{Department of the Physics of Complex Systems, \\
E{\"o}tv{\"o}s University; H-1117 Budapest, Hungary \\
{e-mail: \tt geszti@complex.elte.hu}
}

\begin{abstract}

A double quantum dot charge detector, with one dot Coulomb coupled to the 
electron to be detected and the other modulated by a time-dependent plunger voltage, 
is analyzed in a minimal model. The signal and noise of the detector are calculated by a 
standard  master equation and MacDonald formula technique. We find a dip in the noise 
spectrum at the double Rabi frequency of the double dot, defining the bandwidth available 
for detecting charges in motion. 
\end{abstract} 
\pacs{73.23.Hk, 85.35.Gv, 07.50.Ls}
\maketitle

Mesoscopically narrow passages of electrons, controlling the flow of
electric current by potential barriers just below or above the Fermi
level, are obviously very sensitive to external electric fields,
which suggests their use as charge detectors. Most often, a mesoscopic 
trap or ``island'' carrying the charge is Coulomb coupled to a quantum 
point contact \cite{qpc} or a quantum dot - in that arrangement 
called  a single-electron transistor (SET),\cite{set}, to offer a charge 
sensitivity sufficient to detect individual electrons. Similar tools are 
being explored as readout devices for charge qubits,\cite{qubit} with 
some results emerging about the quantification of the amount of information
accessible in the readout process.\cite{clerk}

Most recently, double quantum dots (DQDs),\cite{vdW} here also called
double island single-electron transistors, have been used for charge 
detection,\cite{brenner} the main advantage being their relative immunity against
noise of various origins. The operation is based on the highly coherent interdot
tunneling, manifest through Rabi oscillations,\cite{hayashi} providing sharp
boundaries on the current-voltage characteristic of DQDs.\cite{liv}. 
This new kind of device has been theoretically analyzed by Tanamoto and 
Hu,\cite{thu} through a model mainly focusing on the complications from weak
Coulomb blocade, and specific aspects related to quantum information, 
also discussed in Ref. \onlinecite{weisslambert}.

In most of the experimental work cited above, the device is operated at
radio frequency, drawing on analogous work on radio-frequency 
SETs,\cite{rfset} efficiently avoiding the band of low frequencies, 
strongly contaminated by $1/f$ noise, which is ubiquitous, and hard
to analyze theoretically. On the contrary, shot noise, extending to much 
higher frequencies, is open to analysis by standard tools, which is also
our main concern here.

\begin{figure}
\includegraphics[scale=0.4]{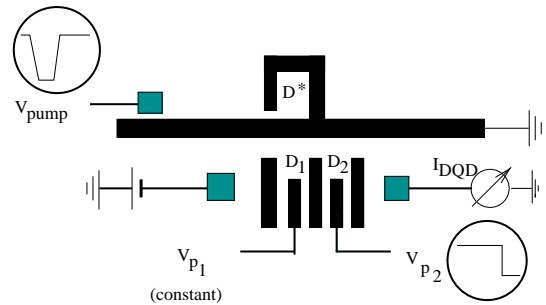} 
\caption{\label{det1.eps}Double quantum dot (DQD) electron detector. The 
 electron to be detected is driven by pump voltage $V_{pump}$ onto the 
 island (trap) $D^*$, then back; its presence or absence is reflected in 
 the current through the DQD\@. The island is Coulomb coupled to compartment 
 $D_1$ of the DQD, set to a fixed plunger voltage $V_{p_1}$; plunger voltage
 $V_{p_2}$ of compartment $D_2$ is used to time-gate the detector.} 
\end{figure}

In the present paper we discuss charge detection in DQDs, supplemented by the 
possibility of fast time control, as sketched in Fig.~\ref{det1.eps}. 
The island trapping the electron to be detected would be placed next to the 
first compartment of the DQD, so the electron would detune the tunneling
resonance set by the two bias plungers.\cite{vdv} Time control, as inspired 
by the experiment of Nakamura {\sl et al},\cite{nak} could be achieved by a 
synchronized modulation of the pump voltage sending the electron onto the 
island, and the {\sl second} plunger of the DQD, driving the device through 
a work point. A variety of delay times and plunger pulse shapes would
allow a detailed analysis of the dynamics of one-electron detection.

In calculating signal and noise of the DQD detector, we evaluate the reduced
density matrix of the coupled DQD+trap system,\cite{nschr} by means of a standard 
Markovian master equation approach (see, e.g., Refs \onlinecite{chenting,kor,gurpra}), 
supplemented by the possibility of resolving the dynamics according to the 
number of achieved tunneling events.\cite{chenting} The latter being slow, 
their fast screening in the external metallic circuits can be approximately 
treated by the Ramo-Shockley theorem in its symmetric form.\cite{RS} 
Frequency-dependent stationary noise is evaluated by means of MacDonald's 
formula,\cite{macd}  which is a shortcut avoiding the more familiar Wiener-Khinchine 
analysis of time correlation functions.\cite{macd1}

The physical time and energy scale of the processes envisaged is set by the Rabi
frequency. For a typical DQD formed by metal contacts on top of a GaAs/AlGaAs 
heterostructure \cite{hayashi} it is of the order $\Omega\approx10^{10}~Hz$. 
To preserve coherence, tunneling rates should be set by gate voltages to the same 
order of magnitude, thereby assuring time control on the $ns$ scale, easy to
follow by external electronics.

Coulomb blockade allows one to consider the low-temperature dynamics of the 
DQD + trap system on a truncated basis, stretched by states of 0 or 1 
extra electron above a neutral background on each dot and the trap. 
A further step of truncation is brought about by the interdot Coulomb repulsion
which for the geometry of the experiment of Hayashi \etal \cite{hayashi} amounts 
to $\approx 10^3~\hbar~\Omega$, effectively excluding states for which both 
dots are occupied.\cite{korpriv,GE}

The above considerations justify the use of a minimal model in which 
the retained basis vectors are denoted as 
$\left|a\right>=\left|000\right>,\,\,\left|b\right>=\left|100\right>,\,\,
\left|c\right>=\left|010\right>,\,\,\left|d\right>=\left|001\right>,\,\,
\left|e\right>=\left|101\right>,\,\,\left|f\right>=\left|011\right>$;
the three occupation numbers belonging to left dot, right dot, and trap respectively. 
In addition, each of those three parts is coupled to a separate metallic contact, acting 
as a fermion reservoir.\cite{trunc} The dynamics of the whole is generated by the Hamiltonian  
${\hat H}={\hat H}_{DQD}+ {\hat H}_T+{\hat H}_{int}$, with 
\begin{subequations}\label{ham}
\begin{align}
&{\hat H}_{DQD}/\hbar=\epsilon_1 a^{\dag}_1a_1 + \epsilon_2a^{\dag}_2a_2 
+\Omega (a^{\dag}_1a_2+a^{\dag}_2a_1)\nonumber\\
&+\sum_lw_{l}b^{\dag}_{l}b_{l} + \sum_rw_{r}b^{\dag}_{r}b_{r}
+\sum_l\lambda_{l}^{*}b^{\dag}_{l}a_1 + h.c. 
\nonumber\\&\hskip4.cm
+\sum_r\lambda_{r}^{*}b^{\dag}_{r}a_2 + h.c.,\\
&{\hat H}_T/\hbar= 
\sum_p w_p d_p^{\dag}d_p+\epsilon_Tc^{\dag}c +
\sum_p\lambda_{p}^{*}d^{\dag}_{p}c + h.c.,\\
&{\hat H}_{int}/\hbar=\nu a^{\dag}_1a_1c^{\dag}c.
\end{align}
\end{subequations}
\n Reduced dynamics of the DQD + trap subsystem are obtained by carrying out thermal 
averaging over the states of the contacts in the initial state.  In the above formulas, 
$\hbar\epsilon_i$ ($i=1,2$) is the energy of an electron occupying the single state 
allowed by Coulomb blockade in the $i$th quantum dot, $\hbar\Omega$ is the amplitude 
of tunneling between the two dots, $\hbar\lambda_l$ and $\hbar\lambda_r$ are the 
respective tunneling amplitudes from left reservoir to left dot and from right dot 
to right reservoir; $\hbar\nu$ is the Coulomb interaction matrix element characterizing 
the coupling of the trap to  the first dot, $\hbar\lambda_p$ is the tunneling amplitude 
between pump and trap, and $\hbar\epsilon_T$ is the energy level of the trap; 
$\hbar w_l$ and  $\hbar w_r$ are the one-electron energies in the left and right contacts, 
and $\hbar w_p$ in the pump. $a_i$ is the one-dot electron annihilation operator, 
$b_l$ and $b_r$ are those for the respective contacts,  $c$ is that of the electron trap, 
and $d_p$ that in the $p$th one-electron state of the pump.

The one-dot levels $\epsilon_i$ can be modulated in time through plunger voltages 
$V_{p_i}$ to scan near-resonance conditions.   Introducing the notation 
$\delta=\epsilon_1-\epsilon_2$,
\be
\delta+\nu=0
\label{res}
\ee
\n is the condition of resonance, assuring that if the trap is occupied by one electron,
the DQD detector exhibits maximum dc transmission, i.e.,  maximum signal with respect
to the empty trap case.

As mentioned above, in conjunction with a Ramo-Shockley treatment of screening,\cite{RS} 
it is convenient to decompose the density matrix according to the total number $N$ of 
tunneling events occurring at both external contacts of the detector, which 
immediately furnishes the Ramo-Shockley screened current as \cite{chenting}
\be
\frac{1}{e}I(t)=\frac12\dot{\bar N}=\frac12\sum_N N\dot p_N(t),
\label{current}
\ee
where
\be
p_N(t)=\sum_i\rho_{ii}^{[N]}(t)=
{\underline v}\cdot\left[{\underline x}_N(t)+{\underline y}_N(t)\right]
\ee
with the notations 
\begin{subequations}\label{vecrho}
\begin{align}
&{\underline x}_N\equiv(\rho_{aa}^{[N]},\,\rho_{bb}^{[N]},\,\rho_{bc}^{[N]},\,
\rho_{cb}^{[N]},\,\rho_{cc}^{[N]})^T,\\
&{\underline y}_N\equiv(\rho_{dd}^{[N]},\,\rho_{ee}^{[N]},\,\rho_{ef}^{[N]},\,
\rho_{fe}^{[N]},\,\rho_{ff}^{[N]})^T,
\end{align}
\end{subequations}                    
and
\be
{\underline v}=(1,\,1,\,0,\,0,\,1)^T.
\ee
The vectors ${\underline x}_N(t)$ and ${\underline y}_N(t)$ are determined by 
the solution of a system of master equations, with the appropriate
initial conditions, which are by no means trivial.\cite{korotkov}

We restrict ourselves to the zero-temperature case; then the Fermi levels of the 
contacts, which control the detector and the pumping of the trap, respectively, do not 
appear explicitly in the calculation, apart from distinguishing pumping-in and 
pumping-out periods, according to whether the pump electrode Fermi level is above or 
below the trap level $\hbar\epsilon_T$. Starting with Hamiltonian (\ref{ham}), and 
following any of the equivalent standard procedures used, {\it e.g.},\/ in the papers 
listed under Refs. \onlinecite{chenting,kor,gurpra}, one invariably arrives at a 
system of Markovian master equations for the components of the density matrix, 
which can be written in the form
\begin{subequations}\label{master}
\begin{align}
&\dot{\underline x}_N={\bf A}\,{\underline x}_N\,+\,{\bf B}\,{\underline x}_{N-1}
+\Gamma_T\,\,{\underline z}_N,\\
&\dot{\underline y}_N={\bf (A+C)}\,{\underline y}_N\,+\,{\bf B}\,{\underline y}_{N-1}
-\Gamma_T\,\,{\underline z}_N\,
\end{align}
\end{subequations}
where
\be
{\underline z}_N=\left\{ \begin{array}{ll} 
                         -{\underline x}_N & ~~\mbox{for pumping-in,}\\
                        ~~{\underline y}_N  & ~~\mbox{for pumping-out,} \end{array} \right.
\ee
and we have introduced the matrices
\begin{subequations}\label{abc}
\begin{align}
{\bf A}&=\begin{pmatrix}
        -\Gamma_L&0&0&0&0\\
        0&0&i\Omega&-i\Omega&0\\
        0&i\Omega&-i\delta-\Gamma_R/2&0&-i\Omega\\
        0&-i\Omega&0&i\delta-\Gamma_R/2&i\Omega\\
        0&0&-i\Omega&i\Omega&-\Gamma_R
        \end{pmatrix},\\
{\bf B}&=\begin{pmatrix}
0&0&0&0&\Gamma_R\\
\Gamma_L&0&0&0&0\\
0&0&0&0&0\\
0&0&0&0&0\\
0&0&0&0&0
        \end{pmatrix},\,\,\,
{\bf C}=\begin{pmatrix}
0&0&0&0&0\\
0&0&0&0&0\\
0&0&-i\nu&0&0\\
0&0&0&i\nu&0\\
0&0&0&0&0
        \end{pmatrix}\,.
\end{align}
\end{subequations}
Here $\Gamma_L$ and $\Gamma_R$ are the familiar Fermi golden rule rates of 
dissipative tunneling transitions from left contact to left dot and from right dot
to right contact, respectively;\cite{truncdamp} $\Gamma_T$ is the rate of transitions 
from pumping contact to trap and vice versa. Damping and dephasing of coherent processes 
in the DQD detector are controlled by these tunneling rates, microscopically rather 
ill defined because of their high sensitivity to defects and impurities, 
but - as already mentioned - easy to control experimentally by gate voltages.

Trivially, the same system of master equations is obeyed by the vectors 
$\underline x,\underline y$ composed of the unresolved density matrix elements 
$\rho_{ij}=\sum_N\rho_{ij}^{[N]}$. It is easy to find their steady-state solution 
$\rho_{ij}^{stac}$; then through Eq. (\ref{current}) we get the steady-state current
\cite{noscrn}  
\be
\frac{I_{stac}}{e}=\frac{\Gamma/3}{\dfrac{1+\alpha/3}{1-\alpha^2}+
\dfrac{1-\alpha}{12}\left(\dfrac{\Gamma}{\Omega}\right)^2
+\dfrac{1/3}{1-\alpha}\left(\dfrac{\tilde\delta}{\Omega}\right)^2}.
\label{istac}
\ee
where 
\be
\tilde\delta=\left\{ 
                     \begin{array}{ll} 
                         \delta & ~~\mbox{for empty trap,}\\
                         \delta+\nu & ~~\mbox{for filled trap;}
                     \end{array} \right.
\label{delta}
\ee
in addition, we have introduced the asymmetry parameter 
$\alpha=(\Gamma_L-\Gamma_R)/(\Gamma_L+\Gamma_R)$ and the mean tunneling rate 
$\Gamma=(\Gamma_L+\Gamma_R)/2$.

The zero-frequency signal of the detector is defined by the difference between 
the current with filled and empty trap, the detector being active all the time. 
Large asymmetries (i.e., $\alpha$ close to unity) cut down the current, which is the 
signal of our detector. Although all the subsequent analysis can be carried 
through for arbitrary values of $\alpha$, the gain thereby is negligible, 
therefore in what follows, we restrict ourselves to the 
symmetric case $\alpha=0$.

Noting that $\delta$ is a ``soft'' experimental parameter, easy to
adjust as desired, we observe that the sensitivity of the current to the 
presence or absence of the coupling term $\nu$ essentially depends on whether 
that coupling is strong ($\nu\gg\Delta$) or weak ($\nu\ll\Delta$) with respect
to
\be
\Delta=\sqrt{3\Omega^2+\Gamma^2/4}.
\label{Delta}
\ee
The conclusion for the experimenter is this: to achieve maximum signal, strive at 
strong coupling in the above sense, and set the bias to $\delta=-\nu$. Taking again
the Hayashi \etal $~$experiment \cite{hayashi} for reference, in all practical cases 
we are in the strong-coupling limit; we note that the actual Coulomb coupling 
strongly depends on the geometry and can be modulated by shifting the island.

\begin{figure}
\epsfig{file=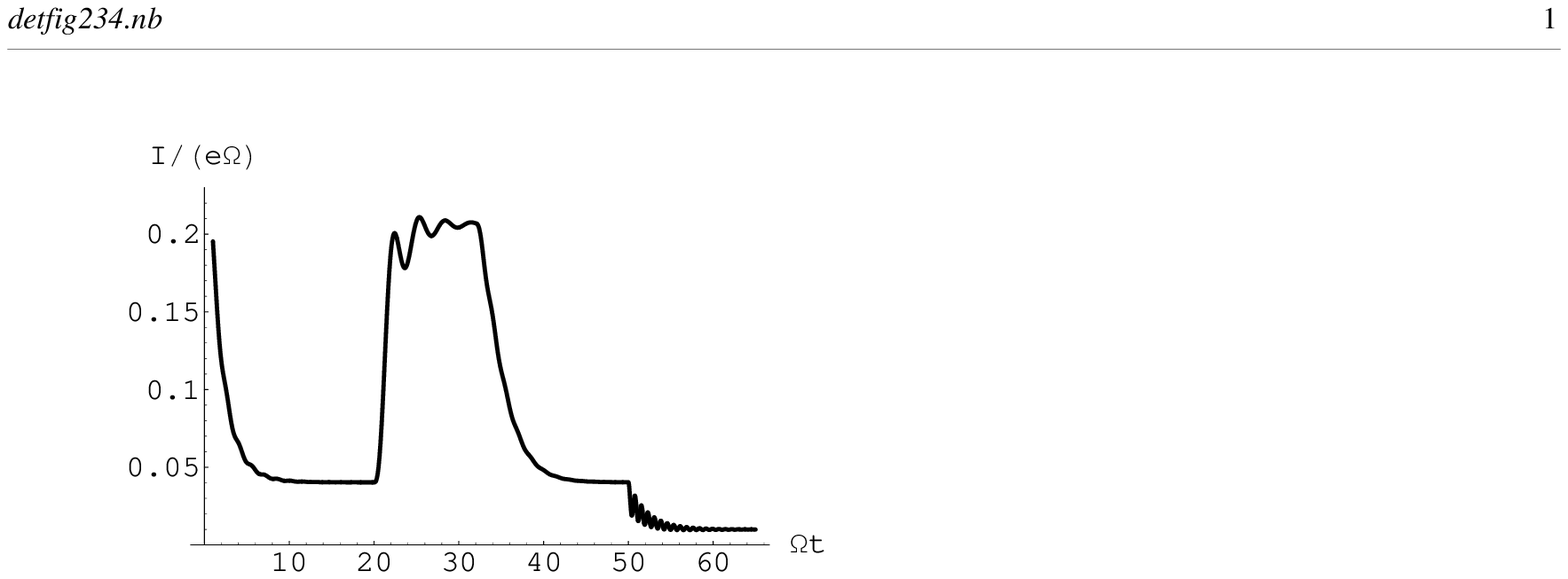,bbllx=85,bblly=600,bburx=350,bbury=695} 
\vskip 0.2cm \hskip 0.3cm
\caption{\label{det2.eps} Current signal of the detector as function of time, 
dimensionless scales being set by interdot tunneling matrix element $\Omega$ and 
electron charge $e$. Parameters $\Gamma/\Omega=2/3,\, \Gamma_T/\Omega=1$ are 
tunable by plunger voltages. For coupling of relative strength $\nu/\Omega=10/3$ 
[moderately strong coupling; see Eq. (\ref{Delta})] which is fixed by 
construction, the DQD is biased to $\delta=-11/3$ to achieve high 
charge sensitivity. Charge is pumped onto the island at $\Omega t=20$ and 
pumped out at $\Omega t=32$; the detector is shut down at $\Omega t=50$ 
by shifting bias $\delta$ to -8.} 
\end{figure}

Detector performance depends on noise as well; in particular, on shot noise, caused 
by the discreteness of electron transport under the joint control of tunneling 
and Coulomb blockade.\cite{blbu} Time-dependent noise analysis, with optimized 
filtering of short and noisy time-gated pulses responding to time-dependent charging
of the trap, is a difficult issue of signal processing. For the present paper, we 
neglect all those complications implicit in Eqs. (\ref{master}), and restrict ourselves 
to the case of constant charge on the trap. Then it is straightforward to calculate the 
frequency-dependent steady-state noise spectrum of the DQD detector, using the now 
standard \cite{mdapp,gb} technique based on MacDonald's formula 
\cite{macd,macd1}
\be
S(\omega)=2 \omega \int_{0}^{\infty} d \tau 
\left(\frac{d}{d \tau}\overline{Q^{2}(\tau)} \right)sin(\omega \tau).
\label{macd1}
\ee
Here $Q(\tau)$ is the total charge carried until time $\tau$,
determined through the time integral of Eq. (\ref{current}), which gives
\be
\frac{d}{dt}\overline{Q^{2}(t)}=\frac{e^2}{4}\sum\limits_{0}^{\infty} 
N^2 \dot{p}_N(t)-2I_{stac}^2 t
\label{macd2}
\ee
in which the long-time linear time dependence of the first term on the right-hand side 
is canceled by the last term. Then the desired noise spectrum is evaluated in the form 
\be
S(\omega)=\omega\frac{e^2}{4i}\left[M(-i\omega)-M(i\omega)\right], 
\label{macd3}
\ee
with
\be
M(z)=\int_{0}^{\infty} dt~e^{-zt}\sum\limits_{0}^{\infty} 
N^2 \dot{p}_N(t),
\ee
to which, as expected, the stationary-current term of Eq. (\ref{macd2}) 
gives no contribution.\cite{nocontr} Accordingly, we proceed through the 
Laplace transform solution of Eqs. (\ref{master}), using the initial 
conditions \cite{korotkov}
\be
\rho_{ij}^{[N]}(0)=\rho_{ij}^{stac}\,\delta_{N0}.
\label{init}
\ee

Having taken the Laplace transform, Eqs. (\ref{master}) furnish an 
iteration in $N$ from $0$ to $\infty$, presenting the Laplace transform 
of Eq. (\ref{macd2}) as a matrix series, which can be summed up in a closed 
form to give
\be
M(z)=z{\underline v}\cdot \frac{{\bf K}+{\bf K}^2}{({\bf I}-{\bf K})^3}\,
\left(z{\bf I}-{\bf A}\right)^{-1}\,{\underline x}_0(0)
\label{2ndmom}
\ee
with ${\bf K}=\left(z{\bf I}-{\bf A}\right)^{-1}\,{\bf B}$, the parameter 
$\delta$ in matrix $\bf A $ [see Eq. (\ref{abc})] being replaced by $\tilde\delta$ 
[Eq. (\ref{delta})] to cover both cases of empty and filled traps. The matrix 
inverses can be evaluated analytically, to finally give the scaled formula
\be
S(\omega\,\vert\,\Gamma,\tilde\delta,\Omega)=
e^2\,\Omega\,\, s\left(\frac{\omega}{\Omega}\,\Big|\,
\frac{\Gamma}{\Omega},\frac{\tilde\delta}{\Omega}\right)
\label{spectrum}
\ee
where $ s(x\,\vert\,y,z)=u(x,y,z)/v(x,y,z),$ with
\be
\begin{split}
 u(x,y,z)&=
4\,y\,\left( 16\,x^8 + 8\,x^6\,
     \left( 7\,y^2 - 4\,\left( 4 + z^2 \right)  \right)  \right. \\& \left.+ 
    x^4\,\left( 57\,y^4 + 16\,{\left( 4 + z^2 \right) }^2 - 
       8\,y^2\,\left( 46 + 11\,z^2 \right)  \right) \right. \\& \left.+
    2\,y^4\,\left( y^4 + 8\,y^2\,\left( -1 + z^2 \right)  + 
       16\,\left( 5 + 6\,z^2 + z^4 \right)  \right)  \right. \\& \left.+ 
    x^2\,y^2\,\left( 19\,y^4 - 
       4\,y^2\,\left( 37 + 10\,z^2 \right)  \right. \right. \\& \left. \left.+ 
       16\,\left( 44 + 23\,z^2 + 3\,z^4 \right)  \right) \right),
\end{split}
\label{u}
\ee
and
\be 
\begin{split}
v(x,y,z)&=16\,x^8 + y^4\,\left(\left( y^2 + 4\,\left( 3 + z^2 \right) 
       \right) \right)^2 + 8\,x^6\,
   \left( 5\,y^2 \right. \\& \left.- 4\,\left( 4 + z^2 \right)  \right) + 
  x^4\,\left( 33\,y^4 + 16\,{\left( 4 + z^2 \right) }^2 - 
     8\,y^2\,\right. \\& \left.\left( 32 + 7\,z^2 \right)  \right)  + 
  2\,x^2\,y^2\,\left( 5\,y^4 + 
     y^2\,\left( 20 - 8\,z^2 \right)  \right. \\& \left.+ 
     16\,\left( 20 + 9\,z^2 + z^4 \right)  \right).
\end{split}
\ee

\begin{figure}
\epsfig{file=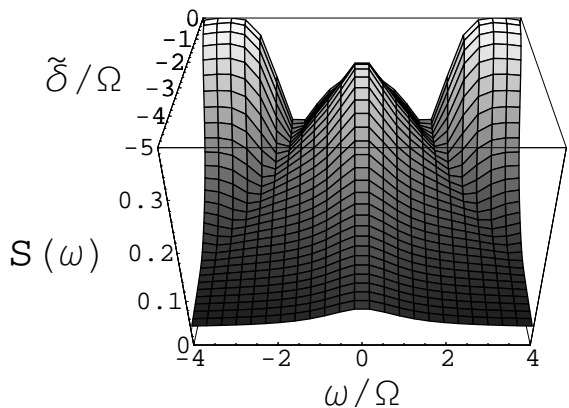,clip=,bbllx=110,bblly=550,bburx=350,bbury=720} 
\vskip 0.2cm \hskip 0.3cm
\caption{\label{det3.eps} Scaled noise spectrum of the double-dot charge 
detector (see text), as a function of relative frequency $\omega/\Omega$ 
and relative bias $\tilde\delta/\Omega$ [see Eq. (\ref{delta})]; with 
relative tunneling rate fixed to $\Gamma/\Omega=1.2$} 
\end{figure}

The results \cite{0} are displayed in Fig.~\ref{det3.eps}. The most conspicuous feature 
is a dip appearing at frequencies $\omega$ slightly below $2\Omega$, the frequency 
of undamped Rabi oscillations. The dip is visible  under near-resonance conditions 
$\tilde\delta\approx 0$, corresponding to peak signal. We think the qualitative reason 
is this. Shot noise is generated when an electron enters or leaves the DQD at an external 
contact, enforcing partition of the current. One oscillation period after entering from 
the left, the electron is just back to the left dot, so it cannot leave right. That kind 
of history is quantified by the solution of the system of master equations (\ref{master}).


In conclusion, our ``minimal model'' offers some insight into the potentialities
of charge detection using a double quantum dot. That refers particularly
to the noise characteristics: 1/f noise being excluded by radio-frequency operation, 
intensive shot noise is expected to appear above the Rabi frequency $2\Omega$, and 
if not cut by the  {\sl RC} time in the external electronics, it can be efficiently 
reduced by low-pass filtering. That sets the intrinsic speed of detection to the 
time scale $\Omega^{-1}$, which in turn depends on geometry and gating voltages; 
according to Ref. \onlinecite{hayashi}, the DQD detector itself can be good 
down to the nanosecond scale. The practical limitation in reaching that 
performance is typically the external circuit.

It is worth mentioning that such arrangements might serve as tools for studying
fundamental issues related to the dynamics of the quantum measurement 
process.\cite{whz,per,giu} Indeed, measuring two-detector correlations between 
detectors differently gated in time may give a new chance to experimentally approach 
the ``collapse of the wave vector,'' which has so far notoriously resisted being 
resolved as a physical process with its own dynamics. Although a disadvantage in 
the qubit readout issue, in this context it may turn into an advantage that the 
operation of fast single-electron detectors is still slower than that of their 
single-photon counterparts, therefore there is more real chance to achieve the 
necessary time resolution for the purpose. Deterministic single-electron sources, 
based on various pump or turnstile constructions\cite{dev} have existed for some 
time, offering the possibility to time-lock detectors to them. What one may happen 
to see is nonstandard precursor fluctuations in the detector current, as the 
random choice inherent in the quantum measurement process develops in time.

We have profited from helpful suggestions by A. N. Korotkov, R. Brenner, and 
S. A. Gurwitz. T.G. would like to thank Moty Heiblum for discussions and 
encouragement offered earlier. This work has been partially supported 
by the Hungarian Research Foundation (OTKA Grants No. T 029544 and 
No. T 049384).


\begin{thebibliography}{99}


\bibitem{qpc}  M. Field, C.G. Smith, M. Pepper, D. A. Ritchie, J. E. F. Frost, 
G. A. C. Jones, and D. G. Hasko, Phys. Rev. Lett. {\bf 70,} 1311 (1993); 
E. Buks, R. Schuster, M. Heiblum, D. Mahalu, and V. Umansky, Nature (London) 
{\bf 391,} 871 (1998); M. H. Devoret and R. J. Schoelkopf, Nature (London) 
{\bf 406,} 1039 (2000).


\bibitem{set}  Yu. Makhlin, G. Sch\"on, and A. Shnirman, Rev. Mod. Phys. 
{\bf 73,}, 357 (2001); N. J. Stone and H. Ahmed, Appl. Phys. Lett. {\bf 77,} 
744 (2000); D. V. Averin, quant-ph/0008004 (unpublished). 


\bibitem{qubit}  A. N. Korotkov, Phys. Rev. B {\bf 60,} 5737 (1999);
S. Pilgram and M. B\"uttiker, Phys. Rev. Lett. {\bf 89,} 200401 (2002);
T. M. Stace and S. D. Barrett, Phys. {\sl ibid.} {\bf 92,} 136802 (2004);
X. Q. Li, W. K. Zhang, P. Cui, J. Shao, Z. Ma, and Y. J. Yan, Phys. Rev. B
{\bf 69,} 085315 (2004); S. A. Gurvitz and G. P. Berman, {\sl ibid.} 
{\bf 72,} 073303 (2005)


\bibitem{clerk} A. A. Clerk, S. M. Girvin, and A. D. Stone, Phys. Rev. B 
{\bf 67,} 165324 (2003).


\bibitem{vdW} W. G. van der Wiel {\sl et al.}, Rev. Mod. Phys. {\bf 75,} 1 
(2003).


\bibitem{brenner} M. Macucci, M. Gattobigio, and G. Iannaccone, 
J. Appl. Phys. {\bf 90,} 6428 (2001); R. Brenner, A.R. Hamilton, 
R.G. Clark, and A.S. Dzurak, Microelectron. Eng. {\bf 67-68,} 826 (2003);
 R. Brenner, A. D. Greentree, and A.R. Hamilton, Appl. Phys. Lett. 
{\bf 83,} 4640 (2003); R. Brenner, T.M. Buehler, and D.J. Reilly, 
J. Appl. Phys. {\bf 97,} 034501 (2005); R. Brenner, T.M. Buehler,
and D. J. Reilly, Microelectron. Eng. {\bf 78-79,} 218 (2005).
Zero-frequency shot noise of the device is analyzed by M. Gattobigio, 
G. Iannaccone, and M. Macucci, Phys. Rev. B {\bf 65,} 115337 (2002).


\bibitem{hayashi} T. Hayashi, T. Fujisawa, H. D. Cheong, Y. H. Jeong, 
and Y. Hirayama, Phys. Rev. Lett. {\bf 91,} 226804 (2003); T. Fujisawa, 
T. Hayashi, H. D. Cheong, Y. H. Jeong, and  Y. Hirayama, Physica E 
(Amsterdam) {\bf 21,} 1046 (2004); J. Gorman, E. G. Emiroglu, D. G. Hasko, 
and D. A. Williams, Phys. Rev. Lett. {\bf 95,} 090502 (2005).


\bibitem{liv}C. Livermore, C. H. Crouch, R. M. Westerwelt,
 K. L. Campman, and A. C. Gossard, Science {\bf 274,} 1332 (1996).


\bibitem{thu} T. Tanamoto and X. Hu, Phys. Rev. B {\bf 69,} 115301 (2004)


\bibitem{weisslambert} S. Weiss, M. Thorwart, and R. Egger, cond-mat/0601699
(unpublished); N. Lambert, R. Aguado, and T. Brandes, cond-mat/0602063
(unpublished).


\bibitem{rfset} R. J. Schoelkopf, P. Wahlgren, A. A. Kozhevnikov, P. Delsing,
and D. E. Prober, Science {\bf 280,} 1238 (1998); M. H. Devoret and 
R. J. Schoelkopf, Nature (London) {\bf 406,} 1039 (2002); A. Aassime, 
D. Gunnarsson, K. Bladh, and P. Delsing, Appl. Phys. Lett. {\bf 79,}, 4031 
(2001).


\bibitem{vdv} N. C. van der Vaart, S. F. Godijn, Y. V. Nazarov, 
C. J. P. M. Harmans, J. E. Mooij, L. W. Molenkamp, and C. T. Foxon, 
Phys. Rev. Lett. {\bf 74,} 4702 (1995).


\bibitem{nak} Y. Nakamura, Yu. A. Pashkin, and J. S. Tsai, Nature (London) 
{\bf 398,} 786 (1999).


\bibitem{nschr} If the trap is part of a coherent charge qubit, through the 
interaction it gets entangled with the charge detector; our theory can be used to
analyze that situation too [J. Zs. Bern\'ad (unpublished)]. Here, as usual, the 
measured object and DQD, as a composite system, are described by standard linear 
quantum mechanics, implicitly assuming that ``wave function reduction'' would take 
place only in the external circuit measuring the current. For a discussion of that 
point, see T. Geszti, Phys. Rev. A {\bf 69,} 032110 (2004). 


\bibitem{chenting}L. Y. Chen and C. S. Ting, Phys. Rev. B {\bf 46,} R4714 
(1992).


\bibitem{kor} A. N. Korotkov, Phys. Rev. B {\bf 49,} 10381 (1994).


\bibitem{gurpra} S. A. Gurvitz and Ya. S. Prager, Phys. Rev. B {\bf 53,} 
15932 (1996).


\bibitem{RS}S. Ramo, Proc. IRE {\bf 27,} 584 (1939); W. Shockley, 
J. Appl. Phys. {\bf 9,} 639 (1938). For a two-contact device, symmetric
screening is the dominant effect; dipolar corrections and other details
related to stray capacitances (Ref. \onlinecite{kor}) are neglected here.


\bibitem{macd} D. K. C. MacDonald, Rep. Prog. Phys. {\bf 12,} 56 (1948).


\bibitem{macd1}  It is worth being mentioned that MacDonald's formula is of full 
quantum validity, utilizing the mean square carried charge, which is determined 
by the {\em symmetrized} \/ current correlation function, just like the 
power spectrum measured by a phase-insensitive device.


\bibitem{korpriv} The importance of interdot repulsion has been pointed out
to us by A. N. Korotkov (private communication).


\bibitem{GE} Because of the elimination of $|11x\rangle$ states, our results 
slightly differ from those obtained by B. Elattari and S. A. Gurvitz, 
Phys. Lett. A {\bf 292,} 289 (2002).

\bibitem{trunc} We notice that since the truncated basis is not a direct product 
of left and right dot subspaces, the operators $a_1$ and $a^{\dag}_2$ do not commute.

\bibitem{korotkov} A. N. Korotkov, Phys. Rev. B {\bf 63,} 115403 (2001); 
R. Ruskov and A. N. Korotkov, {\sl ibid.} {\bf 67,} 075303 (2003).

\bibitem{truncdamp} It is worth mentioning that by excluding the double-occupancy states 
$\left|110\right>$,  $\left|111\right>$ as intermediate states in our second-order 
perturbation calculation for the damping of $\rho_{bc}^{[N]}$, $\rho_{cb}^{[N]}$, 
$\rho_{ef}^{[N]}$ and $\rho_{fe}^{[N]}$, the possibility of damping 
through virtual tunneling across the left contact is automatically excluded. 

\bibitem{noscrn} The steady-state current contains no contribution from external 
screening.


\bibitem{blbu} Y. M. Blanter and M. B\"uttiker, Phys. Rep. {\bf 336,}
1 (2000).


\bibitem{mdapp} See, e.g., D. Mozyrsky, L. Fedichkin, S. A. Gurvitz, and 
G. P. Berman, Phys. Rev. B {\bf 66,} 161313 (R) (2002); R. Aguado and T. Brandes,
Phys. Rev. Lett. {\bf 92,} 206601 (2004). 


\bibitem{gb} S. A. Gurvitz and G. P. Berman, Phys. Rev. B {\bf 72,} 073303 (2005).


\bibitem{nocontr} The mathematical reason is that the Laplace transform of $t$ 
is $z^{-2}$ with zero residue.


\bibitem{0} The trap-contact tunneling rate $\Gamma_T$ would influence quantum 
phase fluctuations but drops out from stationary current noise. The 
zero-frequency value $S(0)$ of the latter can be obtained more directly by 
evaluating the first and second moments of the distribution $P^{[N]}(t)$ in the 
long-time diffusion limit. 


\bibitem{whz} \textit{Quantum Theory and Measurement}, edited by J. A. Wheeler 
and W. H. Zurek (Princeton University Press, Princeton, NJ, 1983).


\bibitem{per}A. Peres, {\sl Quantum Theory: Concepts and Methods}
\/ (Kluwer, Dordrecht 1993).


\bibitem{giu} D. Giulini {\sl et al.}, {\sl Decoherence and the
 Appearance of a Classical World in Quantum Theory} (Springer, Berlin
 1996).


\bibitem{dev} For a review, see M. H. Devoret, D. Esteve, and C. Urbina,
 in {\sl Advances in Quantum Phenomena,} \/ edited by E. G. Beltrametti and
 J.-M. L{\'e}vy-Leblond (Plenum Press, New York, 1995), p. 65.




\end{thebibliography}
\end{document}